\documentclass[12pt]{iopart}

\usepackage{amsfonts}
\usepackage{amssymb}
\usepackage{graphicx}

\begin{document}
\title{Near integrable systems}
\author{E.~Bogomolny$^1$,  M.~R.~Dennis$^2$, R.~Dubertrand$^3$}
\address{$^1$ CNRS, Universit\'e Paris-Sud, UMR 8626\\
Laboratoire de Physique Th\'eorique et Mod\`eles Statistiques, 91405 Orsay,
France \\
$^2$ H.~H.~Wills Physics Laboratory, Tyndall Avenue, Bristol BS8 1TL, UK\\
$^3$ Department of Mathematics, University of Bristol, Bristol BS8 1TW, UK}
%\date{\today}

\eqnobysec

\begin{abstract} 
A two-dimensional circular quantum billiard with unusual boundary conditions
introduced by Berry and Dennis (\emph{J Phys A} {\bf 41} (2008) 135203,
Ref.~\cite{berry}) is considered in detail.  
It is demonstrated that most of its eigenfunctions are strongly localized and
the corresponding eigenvalues are close to eigenvalues of the circular
billiard with Neumann boundary conditions.  
Deviations from strong localization are also discussed. 
These results agree well with numerical calculations.
   
\end{abstract}
%\maketitle 

\section{Introduction}

It is common to consider a quantum problem as integrable (but not necessarily separable) or chaotic, depending solely on the properties of its classical counterpart. 
For example, the quantum circular billiard is integrable because classical mechanics inside a circle is integrable. 
In such an approach one does not even specify the exact form of boundary conditions, which are indispensable for the existence of discrete spectrum in quantum billiards. 
Often this approach is correct; for example a circular billiard remains integrable for both Dirichlet and Neumann boundary conditions. 
Nevertheless, it is known that this is not always the case. 
A notable example is a rectangular billiard with Dirichlet conditions imposed on part of the boundary, and Neumann conditions on the complimentary part (e.g.~\cite{bogomolny} and references therein). 
Though the rectangular billiard is classically integrable, this version of the quantum problem is neither integrable nor chaotic and, in fact, has many features in common with classically pseudointegrable systems. 
In particular, in \cite{bogomolny} it was shown that the eigenfunctions of such systems have a strong resemblance with integrable eigenfunctions whose quantization gives the positions of the energy levels with reasonably good precision.

Recently, a different system with similar properties was introduced in \cite{berry}. 
The problem consists in finding eigenfunctions and eigenvalues for the Helmholtz equation in polar coordinates \cite{berry}
\begin{equation}
(\Delta +k^2)\Psi(r,\phi)=0
\label{equation}
\end{equation}   
with Robin boundary conditions (mixed boundary conditions) on the circle of radius $R$:
\begin{equation}
\frac{\partial }{\partial r}\Psi(r,\phi)|_{r=R}=Af(\phi)\Psi(R,\phi)\ .
\label{bc}
\end{equation}
When the boundary function $f(\phi)$ is a constant, the problem remains integrable (e.g.~\cite{sieber}). 
New phenomena appear when 
\begin{itemize}
\item  $f(\phi)$ is a smooth function of the polar angle $\phi$, and
\item the prefactor $A$ is not a constant but is proportional to the momentum $k$.
\end{itemize}  
In \cite{berry} the case  
\begin{equation}
f(\phi)=\cos \phi
\label{problem}
\end{equation}
and $A=k$ was briefly considered. 
Fig.~7 of that paper shows an eigenfunction with energy $E=97.206986712$ which
looks like the (slightly shifted) standard Bessel function with azimuthal
quantum number $8$ and radial quantum number $6$.  
The $6^{\mathrm{th}}$ zero of $J_8^{\prime}(x)$ is $x_0=27.8892694,$
corresponding to energy $E_0\equiv x_0^2/8=97.226418653.$  
Compared with the unit mean distance between levels, the difference between these is rather small: $E-E_0=-0.01943$.

The purpose of this paper is to show that this not a coincidence. 
We argue that almost all energy levels of the problem (\ref{equation}), (\ref{bc}), and (\ref{problem}) are close to eigenvalues of the circular billiard with Neumann boundary conditions 
\begin{equation}
J_m^{\prime}(x)=0\ .
\label{neumann_cond}
\end{equation}
The corresponding wavefunctions look, roughly speaking, like the usual Bessel function solutions of (\ref{equation}) but slightly shifted with respect to the circle centre.

The main reason for such behaviour is the strong localization of wave functions. 
Namely, the recurrence relation determining the eigenfunction coefficients (see equation (\ref{rr}) below),  is equivalent to the one-dimensional discrete Schr\"odinger equation with a pseudo-random distribution of on-site energies. 
If these energies were truly random, this problem would correspond to Anderson localization in one dimension \cite{anderson_1,anderson_2,pastur} for which all states are localized. 
Though the energy distribution is deterministic in our problem, it has strong pseudo-random properties, and many of the predictions of localization theory remain valid.

The plan of the paper is the following. 
In Section~\ref{localization} the formal solution of the problem in terms of recurrent relations derived in \cite{berry} is discussed. 
Properties of the pseudo-random energy distribution and its relation with the Lloyd model of the product of random matrices are investigated, and the localization length is discussed.  
Properties of strongly localized states and the construction of a local perturbation theory are treated in Section~\ref{strongly}. 
States which are localized close to the boundaries of the allowed region differ from strongly localized states.  
%Wave functions with small azimuthal number can be described by adiabatic
%approximation studied in Section ???.  
In Section~\ref{continuous}, states with large azimuthal number are shown to be adequately described by a continuous semiclassical approximation, whose approximate quantization condition is derived in Section~\ref{approximation}. 
As for any dynamical model, the construction of the semiclassical trace formula is of interest, and this is done for our model in Section~\ref{trace}. 
We conclude in Section~\ref{conclusion}.
Details of longer calculations are given in the Appendices.

\section{Localization}\label{localization}

The solution of the above problem may be written as the following formal series  \cite{berry}  
\begin{equation}
\Psi(r,\phi)=\sum_{m=0}^{\infty} \frac{J_m(kr)}{J_m(x)} a_m \left \{ \begin{array}{c} \cos m\phi \\ \sin m\phi \end{array} \right \}
\label{series}
\end{equation}
where $x=kR$ and $\cos m\phi$  (resp.~$\sin m \phi$) are chosen for states symmetric (resp.~antisymmetric)  with respect to the symmetry transformation $\phi\to -\phi$.

The boundary conditions (\ref{bc}) and (\ref{problem}) are fulfilled provided the coefficients $a_m$ obey the recurrence relations \cite{berry}
\begin{equation}
2\rho_m(x) a_m=a_{m+1}+a_{m-1}
\label{rr}
\end{equation}
where
\begin{equation}
\rho_m(x)=\frac{J_m^{\prime}(x)}{J_m(x)}
\label{rho}
\end{equation}
for all $m=0,1,2,\ldots$. 
The initial values  may be chosen as follows \cite{berry}
\begin{equation}
a_0=1,\;a_1=\rho_0(x)\ ,
\label{even}
\end{equation}
for symmetric functions and
\begin{equation}
a_0=0,\;a_1=1 
\label{odd}
\end{equation}
for antisymmetric ones.

For integrable models such as the circular billiard with Neumann boundary conditions, only one term in the series (\ref{series}) is non-zero. 
For chaotic problems like the stadium billiard, all coefficients are non-zero, and in the mean the numbers $a_m/J_m(x)$ can be considered as independent Gaussian random variables \cite{berry_2}. 
Below we demonstrate that for our problem, almost all states are localized. 
This means that one coefficient $a_m$ is much larger than all the others which, roughly speaking, decrease exponentially from the centre of localization.

\begin{figure}
\begin{minipage}{.49\linewidth} 
\begin{center}
\includegraphics[width=.7\linewidth,angle=-90]{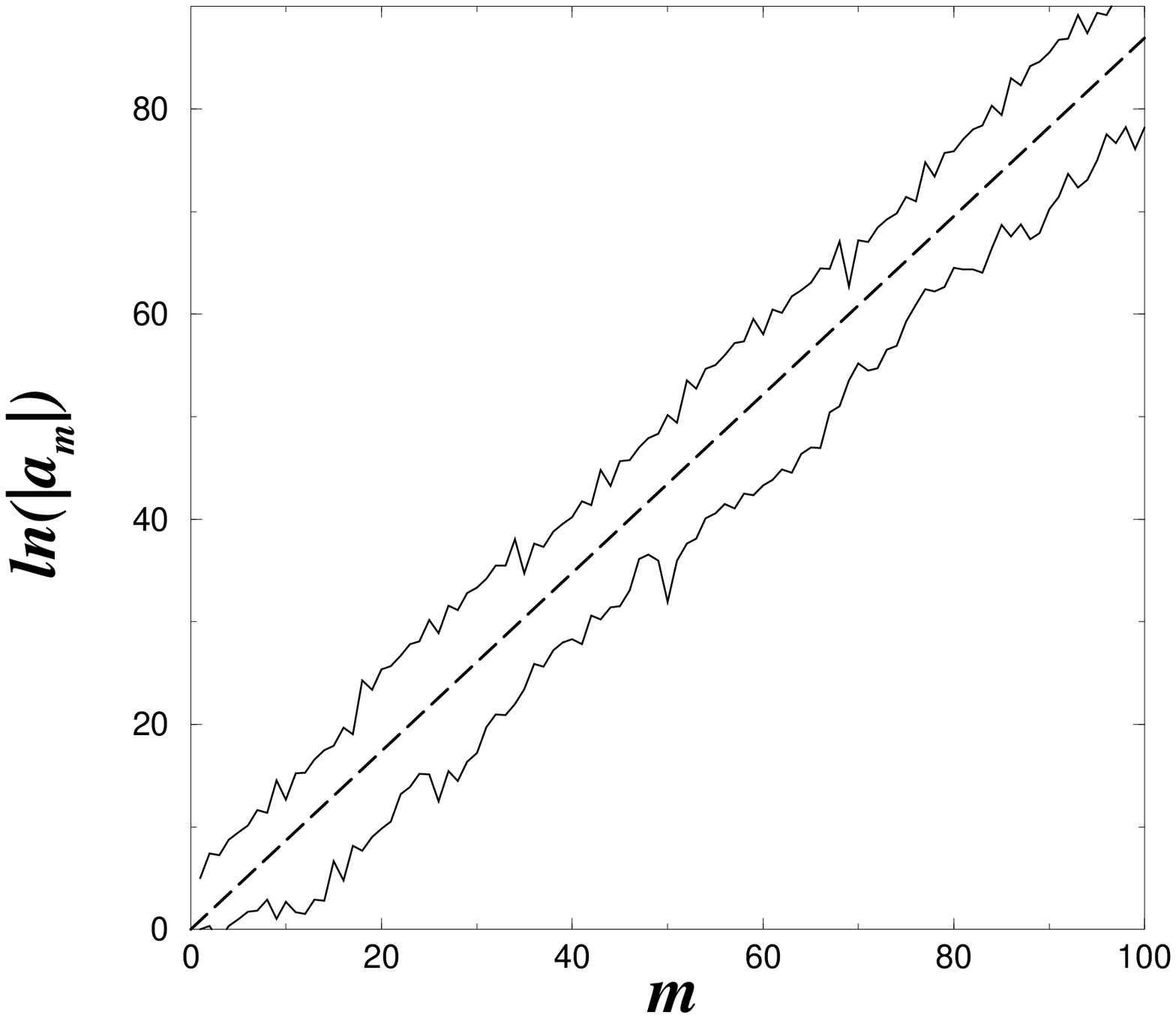}
\end{center}
\begin{center}(a) \end{center}
\end{minipage}\hfill
\begin{minipage}{.49\linewidth} 
\begin{center}
\includegraphics[width=.7\linewidth,angle=-90]{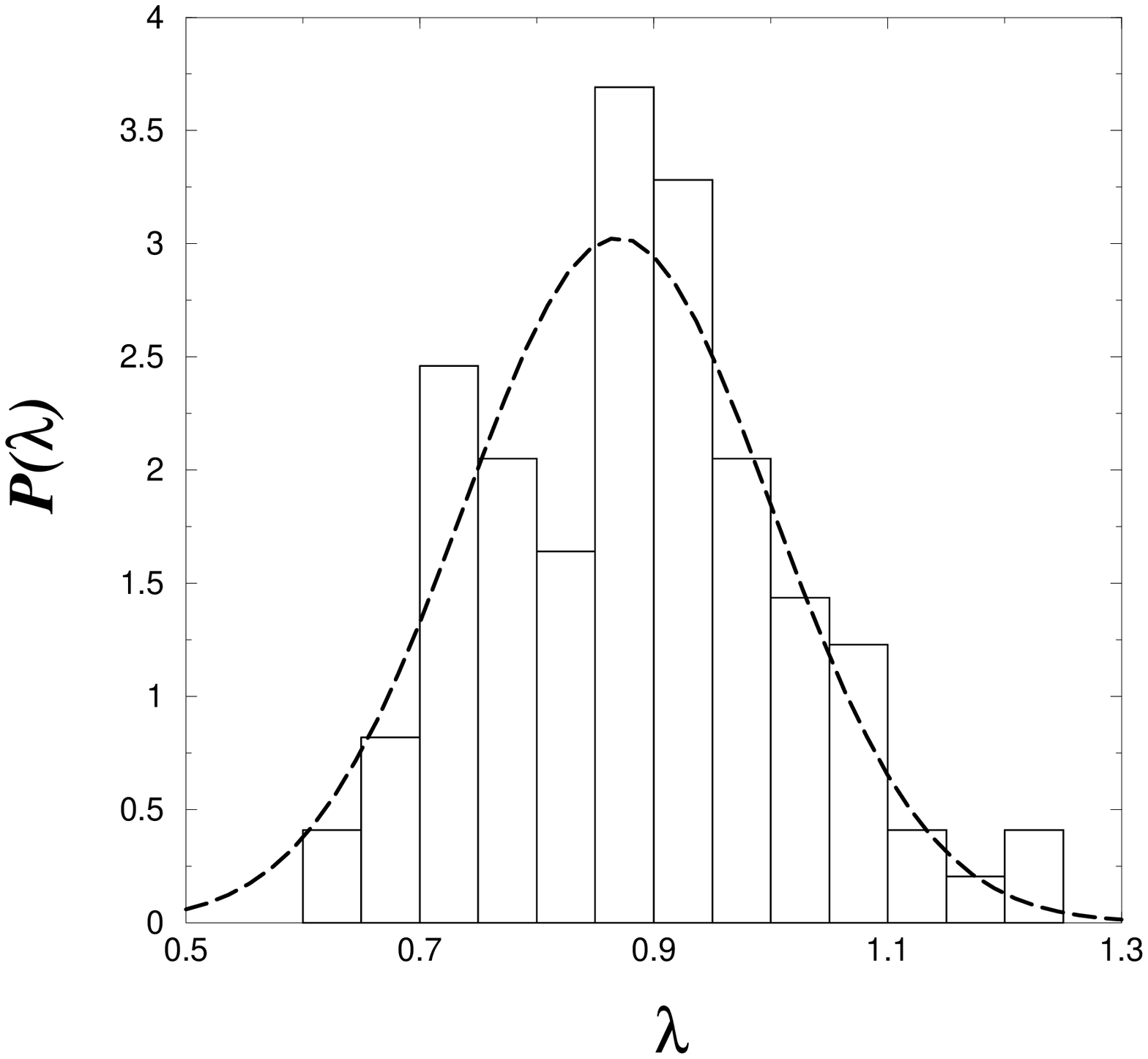}
\end{center}
\begin{center}(b) \end{center}
\end{minipage}
\caption{(a) Logarithm of modulus of the coefficients $a_m$ versus $m$ for the deterministic model (\ref{per_mod}) with $\theta=\pi\sqrt{5}/4$ (the upper solid line) and for one realization of the Lloyd model (\ref{cauchy}) with $w=2\sin \theta$ (the lower solid line). 
For clarity the upper curve is shifted by 5 units. 
The dashed line is the Lloyd model prediction $y=\bar{\lambda} m$ where  $\bar{\lambda}$ is the Lyapunov exponent given by (\ref{lyapunov}). 
(b) Distribution of the Lyapunov exponent for 100 realizations of the Lloyd model with the length $L=100$ and parameters the same as in (a).  
The dashed line is the Gaussian distribution with parameters (\ref{lyapunov}) and (\ref{sigma})}
\label{fig1}
\end{figure}

It is evident that the recursion relations (\ref{rr}) define the discrete Schr\"odinger equation, and can be rewritten in the form of the transfer matrix 
\begin{equation}
\left (\begin{array}{c}a_{m+1}\\a_m\end{array}\right )=\left (\begin{array}{c c} 2\rho_m & -1\\1 & 0 \end{array}\right )  \left (\begin{array}{c}a_{m}\\a_{m-1}\end{array}\right )\ .
\label{product}
\end{equation} 
In the semiclassical approximation $x\to \infty$ and $m<x,$ $\rho_m$ can be approximated by using the standard asymptotics of the Bessel functions \cite{bateman}
\begin{equation}
J_m(x)\approx \sqrt{\frac{2}{\pi}}(x^2-m^2)^{-1/4}\cos \Phi_m(x)\ ,
\label{bessel}
\end{equation}
where 
\begin{equation}
\Phi_m(x)=\sqrt{x^2-m^2}-m\arccos \frac{m}{x} -\frac{\pi}{4}\ .
\label{phase}
\end{equation}
Thus
\begin{equation}
\rho_m(x)\approx -\sqrt{1-\frac{m^2}{x^2}}\tan(\Phi_m(x))\ .
\end{equation}
If, for all $m$, $\Phi_m(x)$ are independent random variables distributed uniformly between $0$ and $\pi$, the variables $\rho_m(x)$ are independent random variables with the Cauchy distribution
\begin{equation}
P(\rho)=\frac{w}{\pi(\rho^2+w^2)}\ ,
\label{cauchy}
\end{equation}  
with
\begin{equation}
w=\sqrt{1-m^2/x^2}\ .
\label{width}
\end{equation}
In such a case, Eq.~(\ref{product}) determines the soluble Lloyd model \cite{lloyd} of the product of random matrices, for which it is known that for almost all initial conditions the Lyapunov exponent of the product, defined
\begin{equation}
\lambda=\lim_{m\to \infty}\frac{\ln \sqrt{a_m^2+a_{m-1}^2} }{m}\ , 
\label{lambda}
\end{equation} 
is non-zero, and for the distribution (\ref{cauchy}) its value is given by \cite{ishii}
\begin{equation}
\bar{\lambda}=\ln{(w+\sqrt{1+w^2})}\ .
\label{lyapunov}
\end{equation}
In parallel with the Lloyd model we also consider a closely related deterministic model (sometimes called the Maryland model \cite{fishman}) defined by the same recursion relations (\ref{rr}), but with the function
\begin{equation}
\rho_m =\sin(\theta)\tan(m\theta)
\label{per_mod}
\end{equation}
with a certain constant $\theta$. 
When $\theta$ is a `good' irrational multiple of $\pi$, this model should have properties close to the Lloyd model. 
In particular, the Lyapunov exponent is given by (\ref{lyapunov}). 
These two models are compared in Fig.~\ref{fig1}a). 
On the other hand, when $\theta/\pi$ is rational (or irrational but with a good rational approximation), the situation differs considerably from a random model (see e.g.~\cite{avron}).

The Lyapunov exponent (\ref{lambda}) is a self-averaged quantity only in the limit $m\to \infty$. 
For a finite sample length (i.e.~large but finite $m\leq L$), it is a random variable with a certain distribution.  
In \cite{boris} it was shown that this distribution in the Lloyd model is Gaussian with mean $\bar{\lambda}$ and variance approximately given by
\begin{equation}
\sigma^2\approx \frac{2\bar{\lambda}}{L}\ .
\label{sigma}
\end{equation}
In Fig.~\ref{fig1}b the distribution of the Lyapunov exponents for the Lloyd model with $L= 100$ is shown. 
For a given realization of a random sequence, the Lyapunov exponent with finite $L$ has a certain value, but the mean value and variance over many realizations agree well with (\ref{lyapunov}) and (\ref{sigma}).

When boundary conditions for the Schr\"odinger equation are imposed at large distances, almost all eigenstates, are localized and for one-dimensional systems the localization length $l$ equals the inverse of the Lyapunov exponent \cite{borland}
\begin{equation}
l=\frac{1}{\lambda}\ .
\end{equation}
For the Cauchy distribution with the width defined by (\ref{width}), the local localization length is
\begin{equation}
l(t)=\left (\ln( \sqrt{1-t^2}+\sqrt{2-t^2}) \right )^{-1}
\label{length}
\end{equation}     
with $t=m/x$ where $m$ is the center of localization and $x$ is the eigen-momentum. 
For a large range of $m/x,$ the localization length is close to $1$ (e.g.~for $m=.5x$, $\bar{l}=1.28$) but near $m=x$ it diverges.

Of course, this statement is valid only for the pure random Lloyd model with the width as in (\ref{width}). 
In our case, the phases $\Phi_m(x)$ in (\ref{phase}) are not random, but rather quickly varying functions of $m$. 
First of all, the oscillatory asymptotics of the Bessel functions (\ref{bessel}) are valid only when
\begin{equation}
|m|<x\ .
\end{equation} 
Therefore, wave functions with eigen-momentum $x$ (if any exist) can be localized only in the interval
\begin{equation}
0<m<x\ .
\end{equation}
Secondly, within this interval, the best one can expect is that the phases (\ref{phase}) are pseudo-random, provided that their derivative over $m$ is not a rational multiple of $\pi,$ nor too close to one. 
This means that in regions where 
\begin{equation}
\frac{\partial }{\partial m}\Phi_m(x)\equiv -\arccos \frac{m}{x}=\frac{M}{N}\pi
\end{equation}    
one cannot expect good localization of eigenfunctions.

Nevertheless, for most values of $m,$ the phases (\ref{phase}) are pseudo-random when considered modulo $\pi$ (cf.~\cite{prange}), and it is natural to assume that wave functions obeying (\ref{rr}) are localized (at least in a certain interval of $m$).

\begin{figure}
\begin{center}
\includegraphics[width=.5\linewidth,angle=-90]{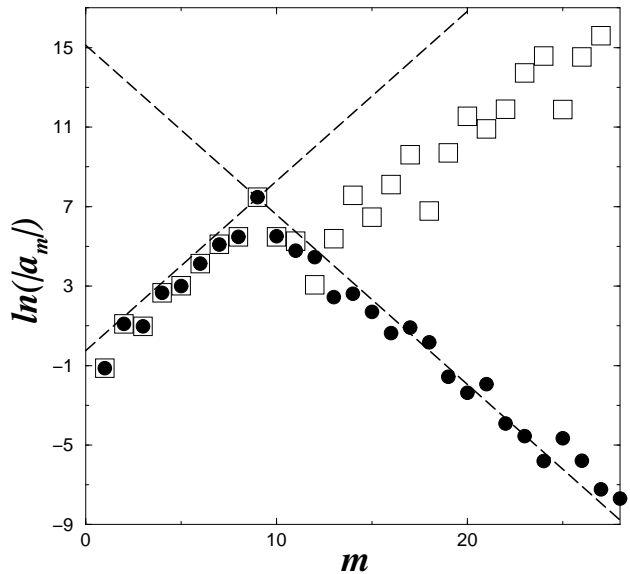}
\end{center}
\caption{Natural logarithm of coefficient modulus $\ln |a_m|$ versus $m$ for a symmetric eigenvalue with $x\approx 32.50302694$ (black circles) and for $x\approx 32.50524735$ corresponding to a zero of $J_9^{\prime}(x)$ (open squares). 
Dashed lines have the slopes given by expression  (\ref{length}) with $m=9$. }
\label{fig4}
\end{figure}

In Fig.~\ref{fig4}, the absolute values of the coefficients $a_m$ corresponding to an eigenvalue are presented on the logarithmic scale. 
It is clear that they correspond to an eigenstate localized at $m=9$ and decaying exponentially from this point with the localization length close to the one given by Eq.~(\ref{length}). 
When the momentum is even slightly different from a true eigenvalue, the coefficients grow exponentially from this point.

\section{Strongly localized states}\label{strongly}

We will call states localized at a certain point $m^*,$  not too close to the boundary $0$ and $x$,  \emph{ strongly localized} states. 
As the localization length $l$ is of order $1$, in general, these states consist of one large component $a_{m^*}$ and all other components have to be small:
\begin{equation}
a_m\sim \mathrm{e}^{-|m-m^*|/l}\ .
\end{equation} 
Therefore it is natural to develop a perturbation series on the number of amplitudes in the vicinity of $m^*$.

We assume that all coefficients except the one with $m=m^*$ are zero, and $a_{m^*}=1$. 
Then from the recurrence relations (\ref{rr}), it follows that the eigenvalue
$x^*$ has to be a zero of the derivative of the Bessel function with $m=m^*,$ 
\begin{equation}
J_{m^*}^{\prime}(x^*)=0\ ;
\label{newmann}
\end{equation}
The next approximation consists of taking into account terms with $m=m^* \pm 1$, which leads to the $3\times 3$ equation
\begin{equation}
\left (\begin{array}{ccc}-2\rho_{m^*-1}(x)&1&0\\1&-2\rho_{m^*}(x)&1\\0&1&-2\rho_{m^*+1}(x)\end{array}\right )
\left ( \begin{array}{c}a_{-1}\\ 1\\ a_{+1}\end{array}\right )=0\ . 
\label{three}
\end{equation}
Of course, this and the following equations can be solved numerically, but it
is more convenient to perform some calculations beforehand.  
We shall see that solutions of any of these equations lead to $x$ close to $x^*$ from (\ref{newmann}). 
Therefore, one has to know the values of $J_{m^*+k}(x^*)$ and $J_{m^*+k}^{\prime}(x^*)$. 
From standard recursion relations for the Bessel functions \cite{bateman}, 
\begin{equation}
zJ_n^{\prime}(z) \pm nJ_n(z)=\pm zJ_{n\mp 1}(z),
\end{equation} 
it follows that the Bessel functions can be written in the form
\begin{equation}
J_{n +k}(z)=J_{n}(z)R_{k,n}(z)-J_{n-1}(z)R_{k-1,n+1}(z)
\end{equation}
where $R_{k,n}(z)$ represents a certain polynomial of degree $k$ in $1/z$ called Lommel's polynomial \cite{bateman}. 
In particular, when $x$ is a zero of $J_m^{\prime}(x)$, direct calculations give
\begin{eqnarray}
\fl
J_{m+1}(x)&=&J_m(x)\frac{m}{x}\ ,\;J_{m-1}=J_m(x)\frac{m}{x}  \ ,\nonumber\\
\fl
J_{m+1}^{\prime}(x)&=&J_m(x)\left (1-\frac{m(m+1)}{x^2}\right )\ ,\;
J_{m-1}^{\prime}(x)=J_m(x)\left (-1+\frac{m(m-1)}{x^2}\right )\ , 
\end{eqnarray}
and
\begin{eqnarray}
\fl
J_{m+2}(x)&=&J_m(x)\left (\frac{2m(m+1)}{x^2}-1\right )\ ,\;
J_{m-2}(x)=J_m(x)\left (\frac{2m(m-1)}{x^2}-1\right )\ ,\nonumber\\ 
\fl
J_{m+2}^{\prime}(x)&=&J_m(x)\left (\frac{2(m+1)}{x}-\frac{2m(m+1)(m+2)}{x^3}\right )\ ,\\
\fl
J_{m-2}^{\prime}(x)&=&J_m(x)\left (-\frac{2(m-1)}{x}+\frac{2m(m-1)(m-2)}{x^3} \right )\ .\nonumber
\end{eqnarray}
Using the properties of Lommel's polynomials or directly applying asymptotic
formulas for the Bessel functions (as in \ref{app_B}), one finds that
the first terms of semiclassical expansion of $J_{p+k}$, calculated at a zero
$x$ of $J_{p}^{\prime}$, are the following 
\begin{equation}
\fl
J_{p+k}(x)=J_p(x)\left ( T_k(u) +\frac{1}{2x(1-u^2)}\left [ ku
    T_k(u)+(k^2(1-u^2)-1) U_{k-1}(u)\right ]\right )+{\cal O}(x^{-2}) 
\label{tchebichef_1}
\end{equation}
and 
\begin{equation}
\fl
J_{p+k}^{\prime}(x)=J_p(x)\left ( (1-u^2)U_{k-1}(u) -\frac{1}{2x}\left [ ku
    U_{k-1}(u)+k^2 T_{k}(u)\right ]\right )+{\cal O}(x^{-2}) 
\label{tchebichef_2}
\end{equation}
where $u=p/x$ and $T_k(u)$ and $U_k(u)$ are the Chebyshev polynomials of the first and the second kind respectively,
\begin{equation}
T_k(\cos\, \theta)=\cos\, k\theta \ ,\;\;U_k(\cos\, \theta)=\frac{\sin\, (k+1)\theta }{\sin\, \theta}\ .
\end{equation}
From these formulas it follows that 
\begin{equation}
\rho_{m^*+k}(x^*)\stackrel{x^{*} \to \infty}{\longrightarrow}\sin \theta \tan k\theta
\label{rho_odd}
\end{equation}
with $\cos \theta=m^*/x^*$, and the last expression is an odd function of $k$: $\rho_{m^*+k}(x^*)=-\rho_{m^*-k}(x^*)$.
In \ref{app_C} it is shown this property implies that all $(2p+1)\times
(2p+1)$ determinants as in (\ref{three}) vanish in the semiclassical limit,
implying that their zeros are always close to the zero of the central element.

For determinants of small size, the first order correction
\begin{equation}
x=x^*+\frac{\delta x}{x^*}
\end{equation}
can be calculated analytically. 
For $3\times 3$ determinant in (\ref{three}) one obtains that 
\begin{equation}
\delta x=-\frac{u^2}{(1-u^2)(2u^4-4u^2+3)}
\end{equation}  
and for the $5\times 5$ determinant
\begin{equation}
\delta x=-\frac{4(8u^6-12u^4+5u^2+1)}{(1-u^2)(64u^8-288u^6+500u^4-388u^2+115)}\ 
\end{equation}
where
\begin{equation}
u=\frac{m^*}{x^*}\ .
\end{equation}
For example, for the eigenvalue $x=32.50302694$ represented in Fig.~\ref{fig4}, one has the following chain of approximations: 
\begin{itemize}
\item For the determinant $1\times 1$, i.e. for the zero of $J_9^{\prime}(x_1)=0$ one has 
\begin{equation}
x-x_1\approx -0.0022 \ .
\end{equation}
\item For the solutions of $3\times 3$, $5\times 5$, and $7\times 7$ determinants one finds numerically the following approximations
\begin{equation}
x-x_3\approx -0.0013 \ ,\;\;x-x_5\approx -0.00022 \ ,\;\;x-x_7\approx -0.000038 \ . 
\label{approximations}
\end{equation}
\end{itemize}
Such perturbation expansions cannot converge uniformly for two different reasons.  
First, although it is reasonable to expect that eigenvalues obtained from such
small determinants give a good approximation to the true eigenvalue
(cf.~(\ref{approximations})) for strongly localized states, the parameter of
such an expansion is a pure number of the order of $\mathrm{e}^{-\lambda}$
where $\lambda$ is the inverse of the localization length.  
Due to considerable fluctuations of the latter, it is difficult to give a
precise \emph{a priori} bound of the accuracy of such series.  
Second, starting from the center of the localization, the correct boundary
conditions -- as in (\ref{even}) and (\ref{odd}) -- are not taken into
consideration.   
Nevertheless, boundary corrections have to be of the order of $\mathrm{e}^{-\lambda \delta N}$ where $\delta N$ is the distance of the center of the localization to the boundary and can often be ignored for strongly localized states. 
In particular, this leads to the almost degeneracy of even and odd states, which differ only by the boundary values (\ref{even}) and (\ref{odd}), which is well-confirmed by numerics.  
For example, the odd state corresponding to even state in Fig.~\ref{fig4} has the momentum $x=32.50302689$ and the difference between even and odd states equals $5.34\ 10^{-8}$.

The above perturbation approach also gives information about the corresponding wave functions. 
Although for strongly localized states, the coefficients $a_m$ decay exponentially from the localization center, numerically this decrease is not so quick, and the influence of the first corrections are noticeable.

We now estimate these corrections from the $3\times 3$ matrix of (\ref{three}). 
Simple calculations reveal that, in the semiclassical limit, the wave function has the form (in this approximation)
\begin{equation}
\fl
\Psi(r,\phi)\sim J_{m^*}(kr)\mathrm{e}^{\mathrm{i}m^*\phi}+
\alpha\left (J_{m^*+1}(kr)\mathrm{e}^{\mathrm{i}(m^*+1)\phi}-J_{m^*-1}(kr)\mathrm{e}^{\mathrm{i}(m^*-1)\phi}\right ),
\label{alpha}
\end{equation}
where $m^*$ is the position of localization center and the correction $\alpha$ is given by
\begin{equation}
\alpha=\frac{1}{2(1-(m^*/x)^2)}\ .
\end{equation}
The fact that the coefficient in front of the $J_{m^*-1}$ term is the opposite of that of the $J_{m^*+1}$ term is not related to the approximation used. 
From \ref{app_C}, it follows that in the semiclassical limit, when (\ref{rho_odd}) is fulfilled, this will always be the case.

\begin{figure}
\begin{center}
\includegraphics[width=.3\linewidth]{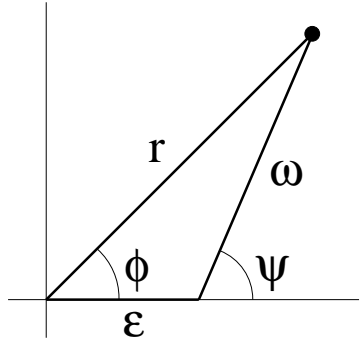}
\end{center}
\caption{Graphical definition of notation for the Bessel function addition theorem (\ref{addition}) }
\label{fig13}
\end{figure}

To see better the main effect of this correction, we consider the familiar Bessel function addition theorem \cite{bateman}. 
According to this theorem, the Bessel function shifted by a vector $\epsilon$ can be expanded as follows:
\begin{equation}
J_m( w)\mathrm{e}^{\mathrm{i}m\psi}=\sum_{n=-\infty}^{\infty}J_n(\epsilon)J_{m+n}( r)\mathrm{e}^{\mathrm{i}(m+n)\phi}
\label{addition}
\end{equation}
where $(r,\phi)$ and $(w,\psi)$ are the polar coordinates of a point with respect to the axis defined by the direction of the shift $\epsilon$ (see  Fig.~\ref{fig13}).

\begin{figure}
\begin{minipage}{.49\linewidth}
\begin{center}
\includegraphics[width=.7\linewidth]{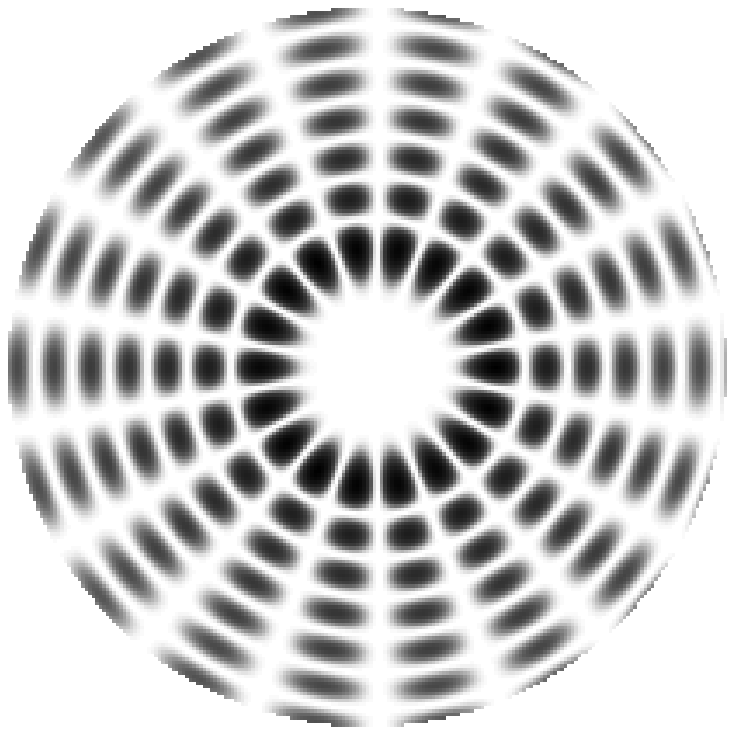}
\end{center}

%\begin{center}a)\end{center}

\end{minipage}\hfill
\begin{minipage}{.49\linewidth}
\begin{center}
\includegraphics[angle=-90,width=.7\linewidth]{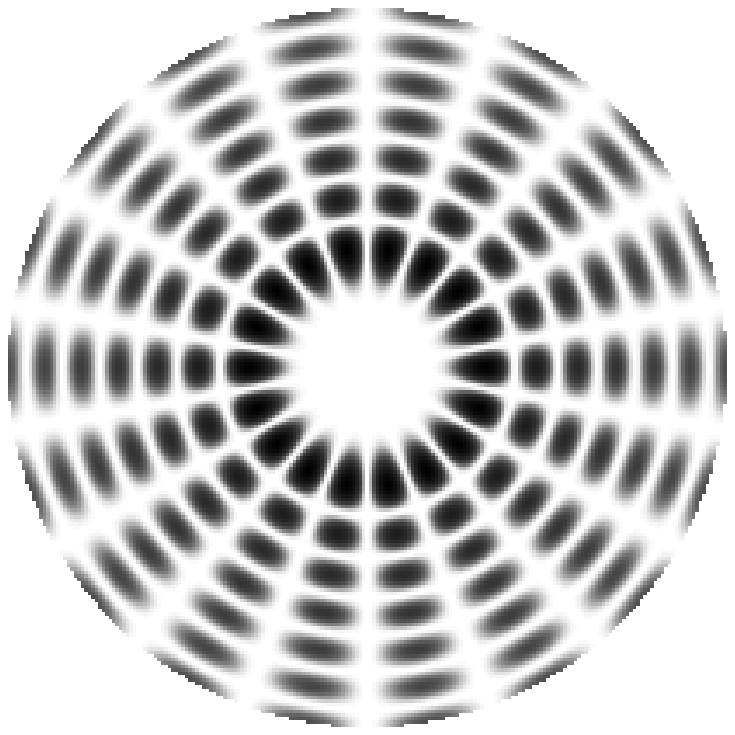}
\end{center}

%\begin{center}b)\end{center}
\end{minipage}

\vspace{2.3cm}

\caption{Left: Wave function of the even state with $x\approx 32.50303$ shown in Fig.~\ref{fig4}. 
Right: Wave function $\Psi(r,\phi)=J_9(kr)\cos 9\phi$ with $kR\approx 32.50525$ satisfying to Neumann boundary conditions.}
\label{fig14}
\end{figure}

Taking into account only the smallest of $\epsilon$ terms, and using $J_{-n}(x)=(-1)^nJ_n(x)$, one concludes that  
\begin{equation}
\fl
J_m(k w)\mathrm{e}^{\mathrm{i}m\psi}\approx  J_0(k\epsilon)\left [ J_{m}( kr)\mathrm{e}^{\mathrm{i}m\phi}+
\beta(J_{m+1}(kr)\mathrm{e}^{\mathrm{i}(m+1)\phi}-J_{m-1}(kr)\mathrm{e}^{\mathrm{i}(m-1)\phi})\right ]
\end{equation}
where
\begin{equation}
\beta=\frac{J_1(k\epsilon)}{J_0(k\epsilon)}\ .
\end{equation}
Comparing this equation with (\ref{alpha}) we conclude that the most noticeable effect of  corrections is a small shift to the left of the state with $m=m^*$, the value of this shift being  of the order of $1/k$. 
For illustration we present in Fig.~\ref{fig14} the numerically-computed wave function corresponding to the eigenstate shown in Fig.~\ref{fig4}.

\section{Continuous approximation}\label{continuous}

\begin{figure}
\begin{center}
\includegraphics[width=.5\linewidth,angle=-90]{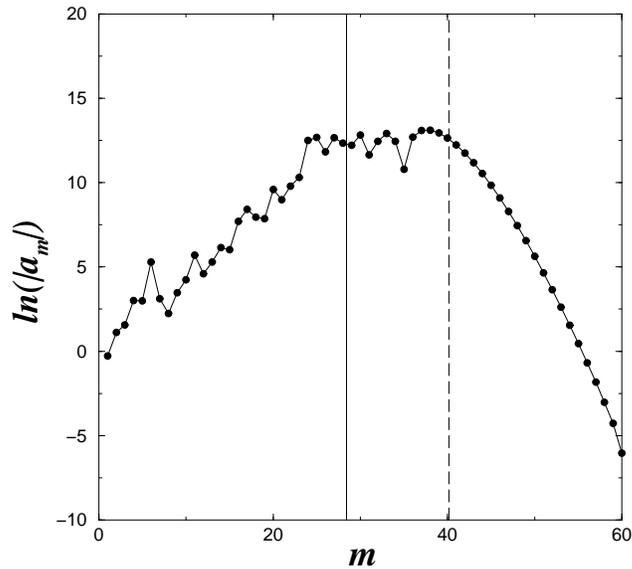}
\end{center}
\caption{Coefficients for an even state with $x=28.40353635$ (black circles connected by solid line). 
Vertical solid line indicates the position of eigen momentum $x$. 
The abscissa of the vertical dashed  line equals $\sqrt{2}x$. }
\label{fig9}
\end{figure}

As the localization length grows when the localization centre $m^*$ is close
to $x$, the structure of the wave functions in this region differs
considerably from the that discussed in the preceding Sections.  
An example is shown in Fig.~\ref{fig9}. 
The absence of a sharp localized peak together with the existence of a flat
part between two vertical lines are a characteristic feature of such a state,
clearly differing from the strongly localized state of Fig.~\ref{fig4}.  
Nevertheless, its eigen-momentum $x=28.40353635$ is still close to the lowest
zero of $J_{26}^{\prime}(x_0),$ at $x_0=28.4181$, the difference being $x-x_0=
-0.0145$. 

\begin{figure}
\begin{center}
\includegraphics[width=.5\linewidth,angle=-90]{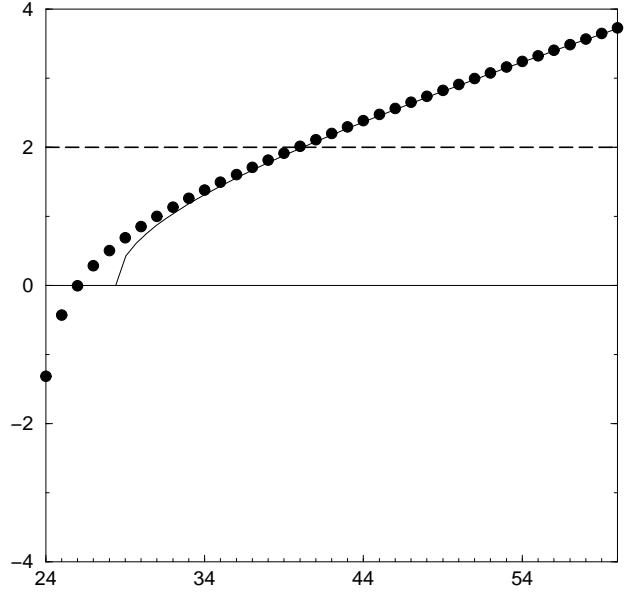}
\end{center}
\caption{`Potential' $2\rho_m(x),$ with $\rho_m(x)$ given by (\ref{rho}) calculated at $x= 28.41807$ as a function of $m$. 
The solid line is the semiclassical approximation (\ref{semi}), defined only when $m>x$. 
The dashed line indicates the value of energy in the continuous Schr\"odinger equation (\ref{cont_equation}). }
\label{fig10}
\end{figure}

To explain such behaviour, we use a semiclassical-type approximation based on
the fact that, in the region $m\approx x,$ the effective `potential' in the
discrete Schr\"odinger equation (\ref{rr}), $V(m)=2\rho_m(x)$, is a smooth
function of $m$ (see Fig.~\ref{fig10}).  
When $m>x$ and $x\to \infty,$ the Bessel function $J_m(x)$ can be approximated
by an asymptotic expression \cite{bateman} similar to (\ref{bessel}): 
\begin{equation}
J_m(x)\approx \frac{1}{\sqrt{2\pi}}(m^2-x^2)^{-1/4}\exp \tilde{\Phi}_m(x),
\end{equation}
where 
\begin{equation}
\tilde{\Phi}_m(x)= \sqrt{m^2-x^2}-m\ln\left (\frac{m}{x}+\sqrt{\frac{m^2}{x^2}-1}\right )\ , 
\end{equation}
from which it follows that  
\begin{equation}
\rho_m(x)\approx \sqrt{\frac{m^2}{x^2}-1}+\frac{x}{2(m^2-x^2)}\ .
\label{semi}
\end{equation}
This formula agrees well with the exact $\rho_m(x)$ for large $m$, except in a region close to $m=x$, where the following uniform approximation is useful (see e.g. \cite{bateman}):
\begin{equation}
\rho_m(x)\approx -\frac{\mathrm{Ai}^{\prime}(y)}{v\mathrm{Ai}(y)},
\label{airy}
\end{equation}
where $\mathrm{Ai}(y)$ is the Airy function of argument
\begin{equation}
y=\frac{1}{v}(m-x)
\end{equation}
and the width $v=(x/2)^{1/3}$.  

In particular, $V(m)$ diverges for $m=m_{p}$, determined by the closest zero of the Bessel function, $J_{m_{p}}(x)=0$. 
For the choice of $x$ in Fig.~\ref{fig10}, $m_{p}\approx 22.79270$. 
In the uniform approximation (\ref{airy}),
\begin{equation}
m_{p}=x-2^{-1/3}\eta x^{1/3}+\mathcal{O}(x^{-1/3})
\label{airy_zero}
\end{equation}
where $\eta$ is the modulus of the first zero of the Airy function, $\eta\approx 2.338$.

We thus have the discrete Schr\"odinger equation
\begin{equation}
V(m)a_m=a_{m+1}+a_{m-1}
\label{discrete}
\end{equation}
with smooth `potential' $V(m)= 2\rho_m(x)$. 
It is therefore natural to look for its particular solution $a_m$ in a
semiclassical form (see e.g. \cite{wkb1})
\begin{equation}
a_m=A(m)\mathrm{e}^{\mathrm{i}\Phi(m)} 
\label{particular}
\end{equation}
with certain smooth functions $\Phi(m)$ and $A(m)$.

From the considerations below it follows that $\Phi(m)\to \infty $ when $x\to \infty$, and $a_{m+1}$ and $a_{m-1}$ may be approximated as the Taylor series  
\begin{equation}
a_{m+1}\approx (A(m)+A^{\prime}(m))\mathrm{e}^{\mathrm{i}[\Phi(m)+\Phi^{\prime}(m)+\frac{1}{2}\Phi^{\prime \prime}(m)]} \ ,
\end{equation}
and
\begin{equation}
a_{m-1}\approx (A(m)-A^{\prime}(m))\mathrm{e}^{\mathrm{i}[\Phi(m)-\Phi^{\prime}(m)+\frac{1}{2}\Phi^{\prime \prime}(m)]} \ . 
\end{equation}
Equating coefficients in the difference equation (\ref{discrete}), in this approximation  
\begin{equation}
\Phi^{\prime}(m)=\arccos \rho_m(x) 
\end{equation}
and 
\begin{equation}
A(m)=\frac{1}{(1-\rho_m^2(x))^{1/4}}\ .
\end{equation}
The oscillating solutions with real $\Phi(m)$ exist when
\begin{equation}
|\rho_m(x)|\leq 1\ 
\label{interval}
\end{equation}
and values of $m$, where $\rho_m(x)=\pm 1$, play the role of turning points. 
Outside the interval (\ref{interval}) one solution is growing and the other is decaying.

Close to the right turning point where $\rho_m(x)\approx 1,$ the continuous Schr\"odinger equation may be constructed directly from the discrete one (\ref{discrete}) by expanding $a_{m\pm 1}$ into a series over the derivatives with respect to $m$:
\begin{equation}
a_{m\pm 1}\approx a_m \pm a_m^{\prime}+\frac{1}{2}a_m^{\prime \prime}\ .
\end{equation}  
One then gets the Schr\"odinger equation
\begin{equation}
\left (\frac{\mathrm{d}^2}{\mathrm{d}m^2}+2 -V(m)\right )\psi(m)=0
\label{cont_equation}
\end{equation}
with 'wave function' $\psi(m)\equiv a_m(x)$.

The solutions (\ref{particular}) are exact analogues of the usual semiclassical solutions of this Schr\"odinger equation, and the matching formulas connecting the decaying solution with the oscillating one may be obtained in the same way as in the standard semiclassical case (see e.g.~\cite{landau}).

We denote
\begin{equation}
q(m)=\left \{\begin{array}{cc}\arccos \rho_m(x)&\;\;\mathrm{when}\;\; |\rho_m(x)|\leq 1 \\
\ln \left (|\rho_m(x)|+\sqrt{\rho_m^2(x)-1}\right )&\;\;\mathrm{when}\;\; |\rho_m(x)|> 1
\end{array}\right. ,
\end{equation}
and let $m_2$ be the right turning point i.e.~the solution of the equation
\begin{equation}
\rho_{m_2}(x)=1 \ .
\end{equation} 
In the semiclassical limit, $m_2$ can be well-estimated from the approximation (\ref{semi}) 
\begin{equation}
m_2=\sqrt{2}x-\frac{1}{2\sqrt{2}}\ .
\label{right_turning}
\end{equation}
When $m>m_2$, the semiclassical solution which tends to zero when $m\to \infty$ has the form 
\begin{equation}
a_m\approx \frac{1}{2(\rho_m^2(x)-1)^{1/4}}\exp \left (-\int_{m_2}^m q(t) \mathrm{d}t\right ),
\label{large}
\end{equation}
and when $m<m_2$, it can be approximated as
\begin{equation}
a_m \approx \frac{1}{(1-\rho_m^2(x))^{1/4}} \cos \left ( \int_{m}^{m_2} q(t) \mathrm{d}t-\frac{\pi}{4} \right )\ .
\label{small}
\end{equation}
When $m>x$ and $m$ is not too close to $x$, one can use semiclassical approximation (\ref{semi}) for $\rho_m(x)$. 
For $m>\sqrt{2}x$ in the leading approximation, one obtains
 \begin{equation}
a_m\approx \frac{1}{2(u^2-2)^{1/4}}\exp \left [-x\int_{\sqrt{2}}^u \ln\left (\sqrt{t^2-1}+\sqrt{t^2-2}\right ) \, \mathrm{d}t\right ],
\label{large_semi}
\end{equation}
and for $x<m<\sqrt{2}x$
\begin{equation}
a_m \approx \frac{1}{(2-u^2)^{1/4}} \cos \left [ x\int_{u}^{\sqrt{2}}\arccos \left (\sqrt{t^2-1}\right ) \,  \mathrm{d}t-\frac{\pi}{4} \right ]\ ,
\label{small_semi}
\end{equation}
where $u=m/x$. 
As expected, the momentum $x$ plays the role of $1/\hbar$.

The left turning point, $m_1$, is defined by the largest solution of 
\begin{equation}
\rho_{m_1}(x)=-1 \ .
\end{equation}
As this point is close to the pole (\ref{airy_zero}), it follows that as $x\to \infty$,
\begin{equation}
m_1\approx m_{p}+1
\label{left_turning}
\end{equation}
where $m_{p}$ is given by (\ref{airy_zero}).

\begin{figure}
\begin{minipage}{.48\linewidth} 
\begin{center}
\includegraphics[width=.9\linewidth,angle=-90]{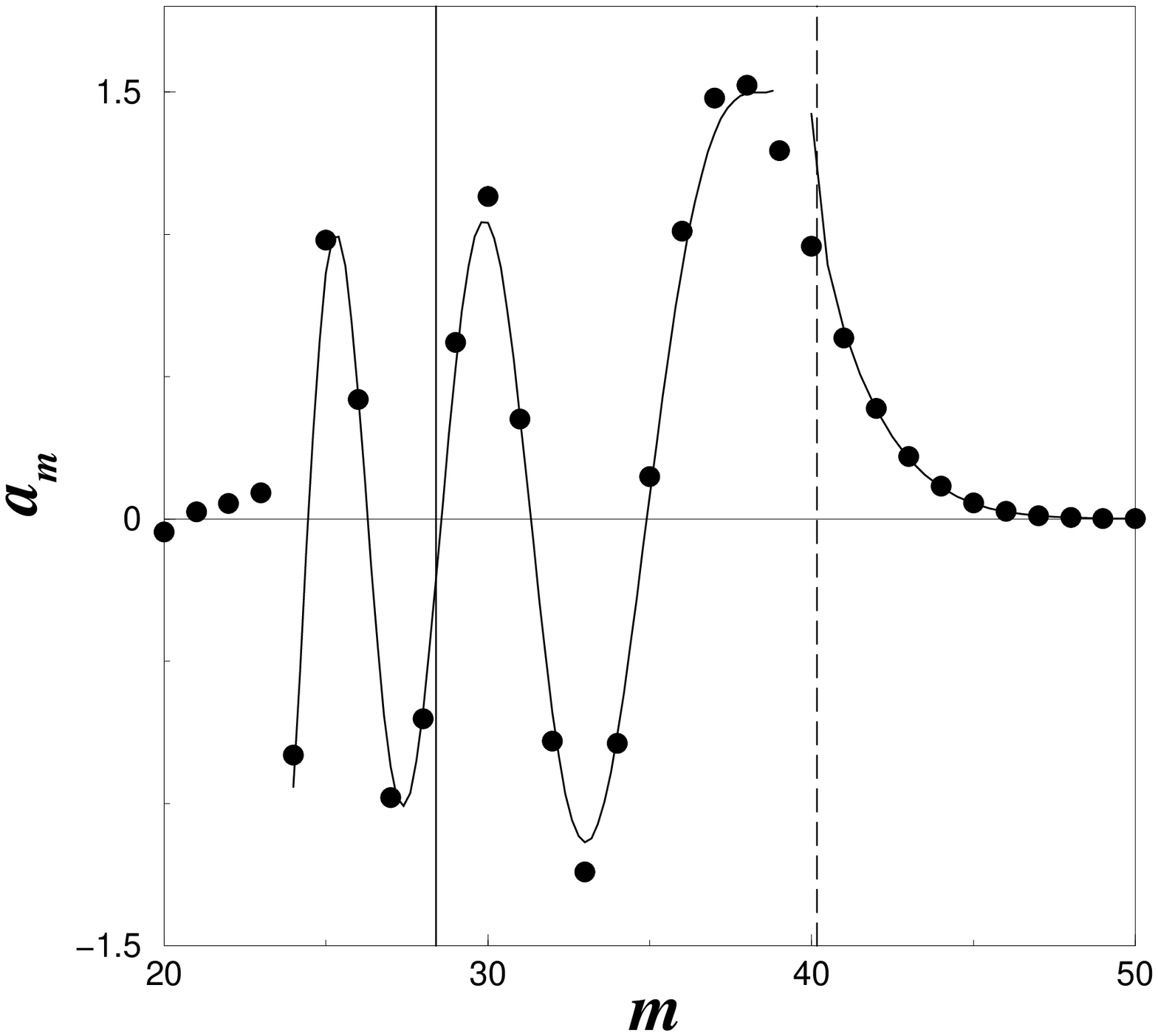}
\end{center}
\begin{center}a)\end{center}
\end{minipage}\hfill
\begin{minipage}{.4\linewidth} 
\begin{center}
\includegraphics[width=.99\linewidth]{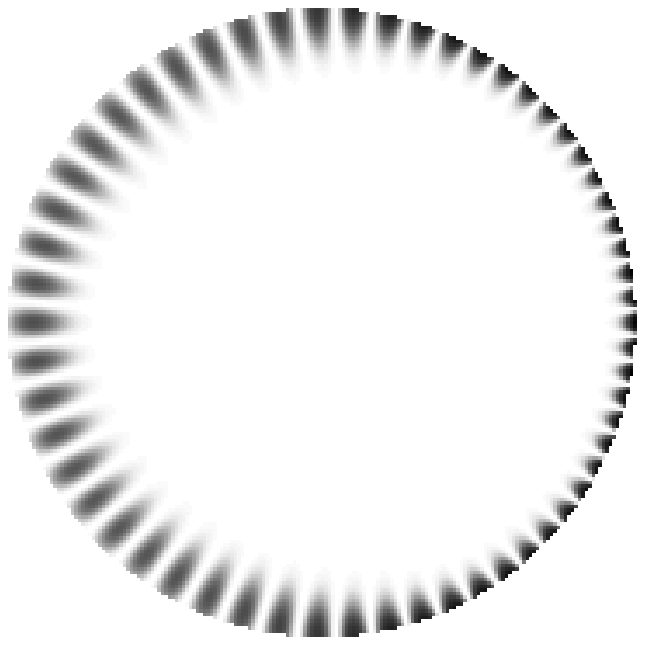}
\end{center}
\begin{center}b)\end{center}
\end{minipage}
\caption{(a) Coefficients $a_m$ for the state of Fig.~\ref{fig9}, normalized to obey the condition (\ref{large}). 
The solid lines are semiclassical formulas (\ref{large}) and (\ref{small}) with $x=x_0$. 
The vertical lines are the same as for Fig.~\ref{fig10}.
(b) Wave function of the state shown in Fig.~\ref{fig9}. } 
\label{fig11}
\end{figure}

Fig.~\ref{fig11}a) shows the result of numerical integration of the semiclassical formulas (\ref{large}) and (\ref{small}) with the exact potential $V(m)$. 
The discontinuity close to $\sqrt{2}x$ is due to the usual inapplicability of semiclassical formulas close to turning points. 
This figure also includes the values of coefficients $a_m$ shown in Fig.~\ref{fig9} on the logarithmic scale. 
To compare them with semiclassical formulas, we normalize the $a_m$ coefficients so that their asymptotics agrees with (\ref{large}). 
To achieve this we multiply all $a_m$ by a factor such that $a_{42}$ equals the value predicted by (\ref{large}). 
The agreement is good , confirming the applicability of the continuous approximation to describe states localized close to $m=x$.  

Though for the states we describe, many coefficients $a_m$ are non-zero, the corresponding wave functions are quite simple (cf. Fig.~\ref{fig11}b)). 
The point is that almost all of these coefficients correspond to the Bessel functions $J_m(x)$ with $m>x$, which decay exponentially inside the circle.

\begin{figure} 
\begin{minipage}{.49\linewidth} 
\begin{center}
\includegraphics[width=.8\linewidth,angle=-90]{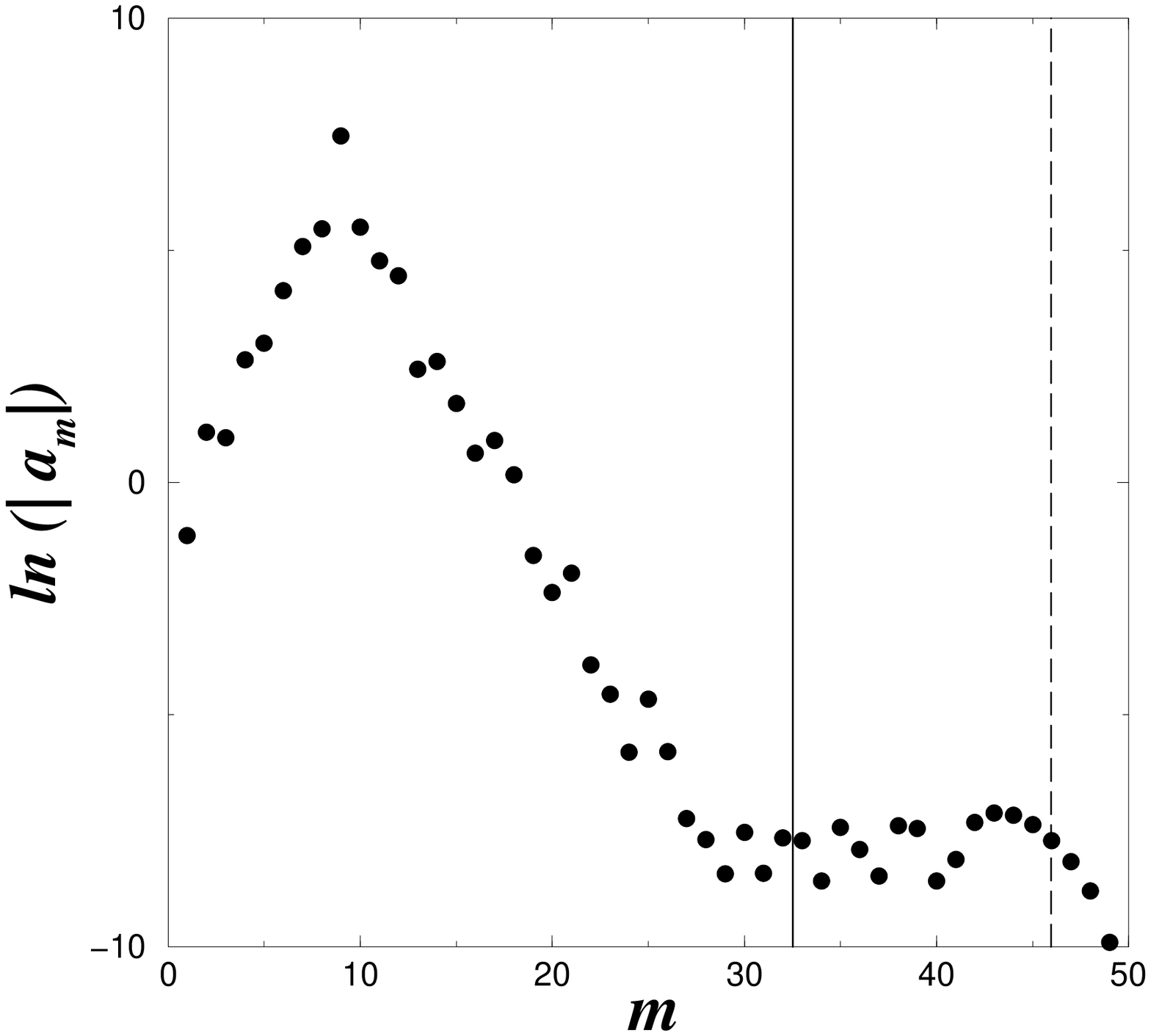}
\end{center}
\begin{center}a)\end{center}
\end{minipage}\hfill
\begin{minipage}{.49\linewidth}
\begin{center}
\includegraphics[width=.8\linewidth,angle=-90]{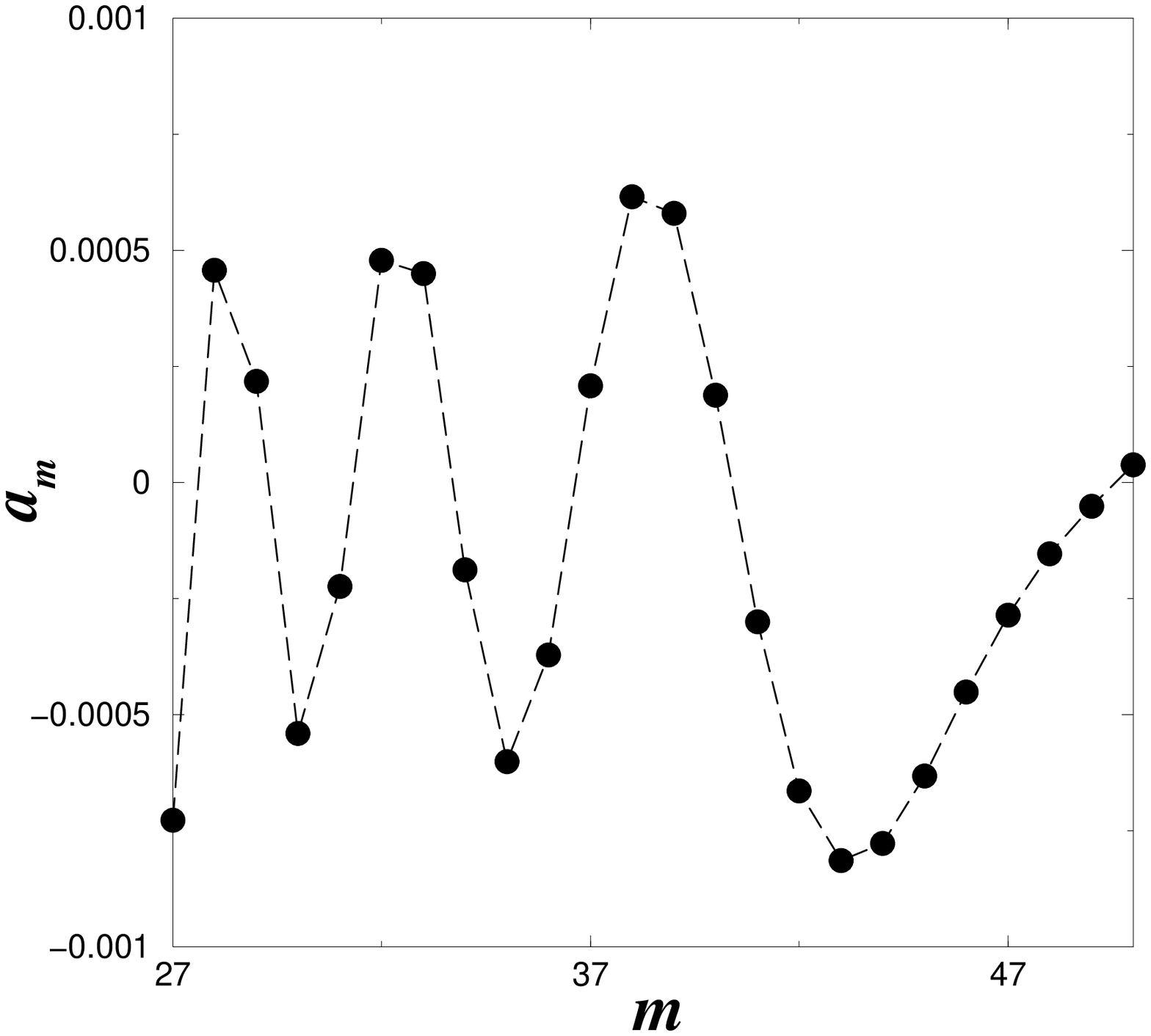}
\end{center}
\begin{center}b)\end{center}
\end{minipage}
\caption{(a) Coefficients $a_m$ on the logarithmic scale for the same state with $x\approx 32.50303$ as in Fig.~\ref{fig4}, but over a larger interval (black circles). 
Vertical solid and dashed lines indicate respectively $m=x$ and $m=x\sqrt{2}$. 
(b) The same as in (a) but in the usual scale and only for high $m$. }  
\label{fig12}
\end{figure}

The behaviour of $a_m$ for large $m$ described in this Section is not, of course, specific to states localized with $m$ close to $x$. 
Rather, it is evidently generic, and all states have a similar form for very large $m$. 
For example, the state in Fig.~\ref{fig4} shows clear exponential localization, but only in the finite interval $m<x$. 
In Fig.~\ref{fig12}a), the same state is plotted over a larger interval. 
Starting from $m=x$, it deviates from pure exponential localization. 
From  Fig.~\ref{fig12}b) it follows that its large-$m$ behaviour is well described by the continuous approximation discussed above.   

\section{Approximate quantization condition}\label{approximation}

If our problem were a problem of a particle in a potential well, the semiclassical quantization condition determining the energy levels inside the well would take the form
\begin{equation}
\int_{m_1}^{m_2}q(m)\,\mathrm{d}m-\frac{\pi}{4}-\delta_1 = \pi p 
\end{equation}
where $\delta_1$ is the phase shift associated with the left turning point and $p$ is an integer.

For the discrete equation (\ref{discrete}), there are two main differences from the standard case. 
First, there are no true local bound states. 
The exact position of the energy levels depends on the precise behaviour of $\rho_m(x)$ far from the left turning point. 
Secondly, in the usual continuous Schr\"odinger equation, the integral of the momentum over the forbidden zone is real, but in our case, between $0$ and $m_1$ it also has an imaginary part equal to $\pi$.

Taking into account this additional phase, in this case the approximate quantization condition has the form
\begin{equation}
\int_{m_1}^{m_2}q(m)\,\mathrm{d}m+\pi\{ m_{p}\}-\delta=\pi p 
\label{quantization_approximate}
\end{equation}
where $m_{p}$ is the position of the pole (\ref{airy_zero}), $\{ f \}$ denotes the fractional part of $f$, and $\delta$ denotes the sum of all phases, which we assume to be a slowly-varying function of the momentum.

As the fractional part of a number differs from that number only by an integer, we can rewrite the above expression as
\begin{equation}
m_{p}+\frac{1}{\pi}\int_{m_1}^{m_2}q(m)\,\mathrm{d}m -\delta= P
\label{quant_1}
\end{equation}
with integer $P$.     

The main term of this approximate quantization condition, when $x\to \infty$, takes the form 
\begin{equation}
x(1+\frac{1}{\pi}I)+g(x)-\delta=P,
\label{quant}
\end{equation}
where $I$ is the elliptic integral (\ref{elliptic})
\begin{equation}
I=\int_1^{\sqrt{2}}\arccos\sqrt{t^2-1}\mathrm{d}t\ .
\end{equation}
$g(x)$ is a function which increases more slowly than $x$. 
A few of its low order terms can be read from (\ref{full_integral}) in \ref{appI},
\begin{equation}
g(x)=-\frac{\eta}{2^{4/3}}x^{1/3}+\frac{1}{6\pi}\ln x +\mathcal{O}(1)\ .
\end{equation}
The states localized in the region close to $x$ are analogues of extreme whispering gallery states. 
For billiards with, say, Neumann boundary conditions, such states are quantized as $x\approx P$ for integer $P$. 
The above discussion demonstrates that for the problem under consideration these states are strongly perturbed, and in the strong semiclassical limit they have momentum
\begin{equation}
x\approx \chi \, P
\end{equation}  
with integer $P$. 
Here the factor 
\begin{equation}
\chi=(1+\frac{1}{\pi}I)^{-1}\approx 0.9\ .
\end{equation}
A physical picture of these states follows.  
Consider the value of the wave function along the boundary of the circle. 
According to (\ref{series}) ,
\begin{equation}
\Psi(R,\phi)=\sum_m a_m\mathrm{e}^{\mathrm{i}m\phi}\ .
\end{equation}
Estimating coefficients $a_m$ from (\ref{small_semi}), one finds
\begin{equation}
\Psi(R,\phi)\sim \sum_m  \cos \left [ \int_{m}^{\sqrt{2}x}\arccos \left (\sqrt{(t/x)^2-1}\right ) \,  \mathrm{d}t-\frac{\pi}{4} \right ]
 \mathrm{e}^{\mathrm{i}m\phi}\ .
\end{equation}
Calculating the sum in the saddle point approximation, we obtain that the saddle point, $m_{\mathrm{sp}}$, obeys the equation
\begin{equation}
\cos \phi=\sqrt{\frac{m_{\mathrm{sp}}^2}{x^2}-1}\ .
\end{equation}
Therefore the saddle point exists if 
\begin{equation} 
\cos \phi>0\ .
\end{equation} 
When this condition is satisfied, the saddle point is the following
\begin{equation}
m_{\mathrm{sp}}=x\sqrt{1+\cos^2 \phi}\ ,
\end{equation} 
and a simple calculation shows that when $-\pi/2<\phi<\pi/2$,
\begin{equation}
\Psi(R,\phi)\sim \exp \left [ \mathrm{i}x\int^{\phi}_0\sqrt{1+\cos^2 f}\mathrm{d}f \right ]\ .
\label{evanescent}
\end{equation}
This wave function corresponds to a local evanescent mode propagating along the boundary 
\begin{equation}
\Psi(r,\phi)\approx \exp \left [ \mathrm{i}x\int^{\phi}_0\sqrt{1+\cos^2 f}\mathrm{d}f -k(R-r)\cos(\phi) \right ] \ . 
\end{equation}
In general, such waves exist when the boundary function $A\,f(\phi)$ entering the boundary conditions (\ref{bc}) is positive \cite{berry}.

Taking into account that at points $\pm \pi/2,$ when $\cos\phi=0$, the angular momentum of wave (\ref{evanescent}) equals $x$, from continuity one concludes that when $\cos\phi$ is negative the wave function along the boundary should have the form 
\begin{equation}
\Psi(R,\phi)\sim \exp \mathrm{i}  x\phi\ .
\label{plane_wave}
\end{equation}  
As the wave function has to be univalued, the quantization condition of such a state is
\begin{equation}
x\left (\pi+\int_{-\pi/2}^{\pi/2}\sqrt{1+\cos^2\phi}\mathrm{d}\phi\right )=2\pi P
\end{equation}
with a certain integer $P$. 
From (\ref{elliptic}) and (\ref{elliptic_2}), it follows that this condition coincides with (\ref{quant}).

\section{Trace formula}\label{trace}

Usual semiclassical arguments lead to the trace formula 
\begin{equation}
d(E)=\bar{d}(E)+d^{\mathrm{osc}}(E),
\end{equation}
where $\bar{d}(E)$ is the smooth part of the level density, and $d^{\mathrm{osc}}(E)$ is its fluctuating part.

From \cite{sieber} and \cite{berry}, it follows that $\bar{d}(E)$ is given by the following Weyl law (for a circle of radius $R$ and $A=k$ in (\ref{bc}) ):    
\begin{equation}
\bar{d}(E)\approx \frac{R^2}{4}+\frac{R}{8\pi k}\int_0^{2\pi}\left [ \frac{2}{\sqrt{1+f^2(\phi)}}-1\right ]\mathrm{d}\phi\ .
\end{equation}
Due to the dependence of $A$ on $k$, the mean part of level counting function, $\bar{N}(E)$, is not just the integral of $\bar{d}(E)$ as usual, but is slightly different :  
\begin{equation}
\bar{N}(E)\approx \frac{1}{4}(kR)^2+\frac{1}{4\pi}kR\int_0^{2\pi}\mathrm{d}\phi\left [2\left (\sqrt{1+f(\phi)^2}-f(\phi)\right )-1\right ]\ .
\end{equation}
The fluctuating part of the level density,  $d^{\mathrm{osc}}(E)$, is given to leading order by the sum over all periodic orbits (see e.g.~\cite{remy}):
\begin{equation}
d^{\mathrm{osc}}(E)=\sum_p\frac{\mathcal{A}_p}{\pi\sqrt{2\pi k l_p}}\langle R_p\rangle \mathrm{e}^{\mathrm{i}[k l_p -\mu_p +\pi/4]} +\mathrm{c.c.},
\label{fluctuating}
\end{equation}
where $l_p$ is the periodic orbit length, $\mathcal{A}_p$ is the area swept by the periodic orbit family, $\mu_p$ is the phase accumulated due to the caustics, $R_p$ is the total reflection coefficient for a given periodic orbit equal the product of reflection coefficients in all points of reflection, and $\langle R_p\rangle$ is its average value over all initial points.

Periodic orbits for the circle are regular polygons characterized by two integers $N$ and $M$. 
The integer $N$ gives the number of reflections with the boundary, and the integer $M$ determines the number of full rotations around the origin. 
For co-prime $N$ and $M$, the periodic orbit is primitive. 
Otherwise, it corresponds to the $r^{\mathrm{th}}$ repetition of a primitive periodic orbit where $r=(M,N)$ is the largest common factor of $M$ and $N$.

\begin{figure}
\begin{center}
\includegraphics[width=.4\linewidth]{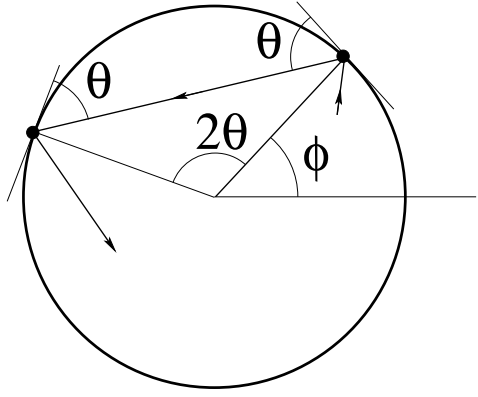}
\end{center}
\caption{Geometrical reflection from the circle. }
\label{fig5}
\end{figure}

For the circle, 
\begin{equation}
l_p=2RN\sin \theta_{M,N}\ ,\;\;
\mathcal{A}_p=\pi R^2\sin^2\theta_{M,N}\ ,\;\;
\mu_p=\frac{\pi}{2}N\
\end{equation}
where 
\begin{equation}
\theta_{M,N}=\pi \frac{M}{N}\ .
\label{theta}
\end{equation}
For the problem under consideration, each reflection with the boundary corresponds to the following reflection coefficient \cite{berry}:
\begin{equation}
R=\frac{\sin \theta-\mathrm{i}\cos \phi}{\sin \theta+\mathrm{i}\cos \phi}
\label{reflection_coefficient}
\end{equation}  
where $\phi$ is the polar angle of the collision point and $\theta$ is the angle between the trajectory and the tangent at the point of incidence (see Fig.~\ref{fig5}). 

For a periodic orbit determined by integers $M,N$, the total reflection coefficient is the product of the reflection coefficients for all points of the collisions
\begin{equation}
R_p(\phi)=\prod_{n=0}^{N-1}\frac{\sin \theta_{M,N}-\mathrm{i}\cos (\phi+2n\theta_{M,N})}{\sin \theta_{M,N}+\mathrm{i}\cos (\phi+2n\theta_{M,N})}
\label{theproduct}
\end{equation}
where $\theta_{M,N}$ is given by (\ref{theta}), and $\phi$ is the polar angle of the initial incident point.

For co-prime $M$ and $N$, this product is calculated analytically in \ref{app_A}, with final result 
\begin{equation}
R_p(\phi)=\left \{ \begin{array}{cl}1&\;\;N=2q\\
\frac{\sinh \Lambda N-\mathrm{i}(-1)^q\cos \phi N}{\sinh \Lambda N+\mathrm{i}(-1)^q\cos \phi N}&\;\;N=2q+1
\end{array}\right . .
\end{equation}
Here, $\Lambda$ is determined from the equation $\sinh \Lambda=\sin \theta_{M,N}$, or
\begin{equation}
\Lambda=\ln \left (\sqrt{1+\sin^2 \theta_{M,N}}+\sin \theta_{M,N}\right )\ .
\end{equation}
For a repetition of a primitive periodic orbit (i.e.~when $M=rM_0$  and $N=rN_0$ with $(M_0,N_0)=1$), the reflection coefficient is  
\begin{equation}
R_{pr}=(R_{p}(\phi))^r
\end{equation}
where $R_{p}(\phi)$ is the reflection coefficient for the primitive orbit determined by $M_0$ and $N_0$.

To calculate the mean reflection coefficient it is necessary to integrate the above formulas over all initial incidence angles $\phi$:
\begin{equation}
\langle R_{pr}\rangle =\int_0^{2\pi}(R_p(\phi))^r\frac{\mathrm{d}\phi}{2\pi}\ .
\label{mean_coefficent}
\end{equation}
This integral can be calculated analytically for all $r$:
\begin{equation}
\int_0^{2\pi}(R_p(\phi))^r\frac{\mathrm{d}\phi}{2\pi}=(-1)^r+\sinh \Lambda N P_{2r-1}\left (\frac{1}{\cosh \Lambda N}\right )
\end{equation}
where $P_{2r-1}(x)$ is a polynomial of degree $2r-1$ of the variable $x=1/\cosh \Lambda N$.

In particular, for $r=1,2,3,4$,
\begin{equation}
\fl
P_1(x)=2x, \;\;P_3(x)=-4x^3, \;\; P_5(x)=12x^5-8x^3+2x, \;\; P_7(x)=-40x^7+48x^5-16x^3.
\end{equation}
For large odd $N$ and fixed $r$, the average reflection coefficient is exponentially close to the Neumann value $1$, 
\begin{equation}
\langle R_p\rangle\approx 1-4r^2\mathrm{e}^{-2\Lambda N}\ .
\end{equation} 
For the triangular periodic orbit with $N=3$ and $M=1$, the average reflection coefficient is also very close to $1$: $\langle R_3\rangle\approx .96391$, but for its repetitions it starts to deviate from it. 
For example, for the second, third, and fourth repetition, it has the following values: $.85972$, $.69842$, and $.49720$.

Each term in the oscillating part of the trace formula (\ref{fluctuating}) is of the order of $k^{-1/2}$, which is the dominant contribution when $k\to \infty$. 
In general, there exist terms decreasing as higher power of $k$. 
Usually, these are just small corrections to the existing periodic orbit amplitudes and are rarely taken into account. 
For example, trace formulas for odd and even states are slightly different due to mainly the existence of rogue states \cite{berry}. 
These eigenfunctions exist only for even states and are described by
\begin{equation}
\Psi(r,\phi)=J_0(k_0r)+\frac{J_0(k_0R)}{J_1^{\prime}(k_0R)}J_1(kr)\cos \phi,
\end{equation}
where $k_0$ is determined from the condition $J_1(k_0R)=0$. 
Since $J_0^{\prime}(x)=-J_1(x)$, these states are the exact analogues of $m=0$ states for pure Neumann boundary conditions and, as it is easy to check, give corrections proportional to $k^{-1}$ to the diameter orbit and its repetitions corresponding to $l_p=2Rj$ with integer $j$. 
Notice that peaks with odd $j$ are absent in the usual trace formula (\ref{fluctuating}) which is typical for desymmetrized systems.

In the problem we consider here the situation is different. 
The shift of the whispering gallery mode (\ref{quant}) produces unusual peaks in the trace formula.  
According to (\ref{quant}), the level density of the new levels are given by the formula
\begin{equation}
\fl
d(x)\equiv\sum_{P=-\infty}^{\infty}\delta(x-x_P)= \sum_{P=-\infty}^{\infty}\left (C+\frac{\mathrm{d}g(x)}{\mathrm{d}x}\right )\delta(P-xC-g(x)+\delta),
\end{equation} 
where 
\begin{equation}
C=1+\frac{1}{\pi}I=\frac{1}{2}+\frac{1}{\pi}\int_0^{\pi/2}\sqrt{1+\cos^2\phi}\mathrm{d}\phi\ .
\end{equation} 
Using the Poisson summation formula we conclude that
\begin{equation}
d(x)=\left (C-\frac{\eta}{32^{4/3}x^{2/3}}+\frac{1}{6\pi x} \right )(1+2\sum_{r=1}^{\infty}\cos(L(x)r))
\end{equation}
with 
\begin{equation}
L(x)=2\pi (xC-\frac{\eta}{2^{4/3}}x^{1/3}+\frac{1}{6\pi}\ln(x)) +\cal{O}(1)\ .
\end{equation}
Therefore in the limit $x\to \infty$ these states correspond to a new periodic orbit with the length equal to
\begin{equation}
l=2\pi R C\approx 6.92 R\ .
\label{additional}
\end{equation}
At finite $x$, the peak associated with this orbit is slightly smaller due to the $x^{1/3}$ correction which dies slowly.

\begin{figure} 
\begin{center}
\includegraphics[width=.7\linewidth]{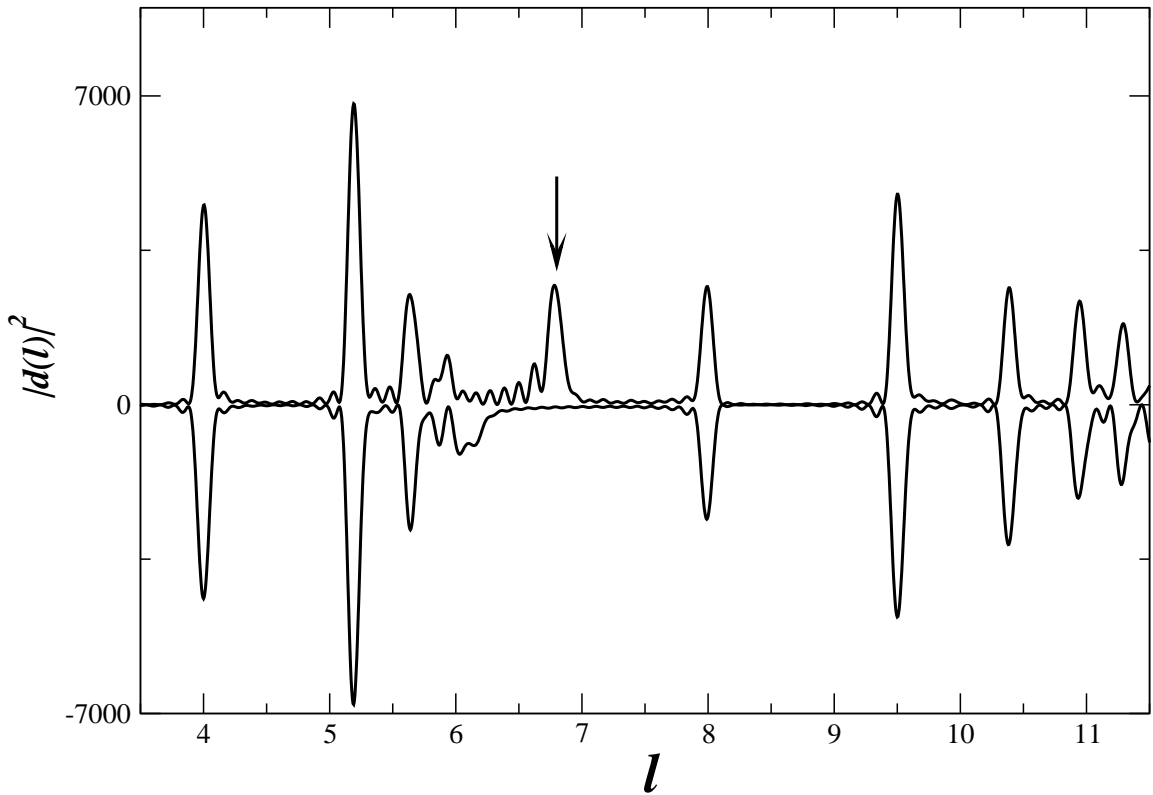}
\end{center}
\caption{Upper curve: oscillating part of the length density. 
Lower curve: the same but for Neumann boundary conditions. 
The arrow indicates additional orbit (\ref{additional}).  }  
\label{fig18}
\end{figure}

The numerically-computed length density of orbits is shown in Fig.~\ref{fig18}. 
For comparison, the length density for the Neumann boundary conditions is also represented in the same figure but, for clarity, with the opposite sign.  
All the peaks coincide with those of the circle except the additional one associated with the orbit (\ref{additional}) indicated by the arrow. 
The analysis of the amplitudes of the peaks (e.g.~for the triangular orbit and its repetitions) confirms  (\ref{mean_coefficent}). 

\section{Conclusion}\label{conclusion}

In this paper, we studied properties of a circular quantum billiard with
specific boundary conditions introduced in \cite{berry}. 
Due to the explicit dependence of the boundary conditions on momentum, the
semiclassical limit of this model is unusual and is neither integrable nor
chaotic.  
Following \cite{berry}, we denote  such systems as near integrable. 
Their characteristic property is the strong localization of wave functions in
the space of azimuthal quantum numbers.  
The main reason of such behaviour in this system is the formal analogy between
the recurrence relations for coefficients of the eigenfunction expansion, and
the one-dimensional Anderson model.  
In a sense, our system is similar to kicked systems where the localization has
been established in \cite{prange}, the role of kicks being played by
collisions with the boundary.

The eigenfunction with momentum $k$ can be localized with azimuthal quantum
number $m$ between $0$ and $kR$.  
States localized far from the boundaries of this interval decay (in the mean)
exponentially from the point of localization with localization length of the
order of $1$.  
Eigen-momenta of these states are close to zeros of the derivative of the
Bessel function $J_m^{\prime}(x)$ with $m$ equal the localization center.

States localized close to the boundaries may deviate from pure exponential
localization.  
In particular, states having large components with $m$ near $kR$ exponentially
decrease only for smaller $m$.   
The large-$m$ behaviour of these states is described by a continuous
approximation, and they have oscillations for $m$ in between $kR$ and
$\sqrt{2}kR$, decreasing quickly only for $m>\sqrt{2}kR$.  
The possibility of formation of fractal states similar to those investigated
in \cite{altshuler} requires additional study.

Among other consequences of strong localization it is worth mentioning the almost-degeneracy of states with different symmetries, and the Poissonian character of spectral statistics for energy eigenvalues with the same symmetry. 

As for usual quantum dynamical systems, it is possible to write down a semiclassical trace formula relating the quantum spectrum with the sum over periodic orbits. 
For the problem considered, all periodic orbits but one are the same as the integrable case of the circular billiard with Neumann boundary conditions, but their amplitudes are different due to a different coefficient of reflection with the boundary. 
The exceptional orbit is related to a partially evanescent mode and its length is unusual (see (\ref{additional}).

Though throughout the paper we focus only on a particular example of boundary conditions (with $f(\phi)=\cos(\phi)$), our discussion is general and the generalization for other boundary functions should be straightforward. 

\section*{Acknowledgements}

MRD is supported by the Royal Society of London. RD wishes to acknowledge
financial support from the ``Programme Lavoisier'' of 
the French Minist\`ere des Affaires \'etrang\`eres et europ\'eennes and from EPSRC.

\appendix
\section{Vanishing of a determinant}\label{app_C}
Consider the $(2p+1)\times (2p+1)$ tri-diagonal matrix whose near-diagonal elements equal $1$, and diagonal elements $V_n$, enumerated consecutively from $-p$ to $p$, obey 
\begin{equation}
V_0=0 \ ,
\label{zero}
\end{equation}
and
\begin{equation}
V_{-n}=-V_n\;\; \mathrm{for }\;\; n=1,\ldots, p\ .
\label{non_zero}
\end{equation}
The purpose of this Appendix is to prove that the determinant of this matrix is zero. 
To do this, we show that this matrix always has an eigenvector, $a_n$, with zero eigenvalue.

By definition $a_{p+1}\equiv 0$, so it is always possible to find quantities $a_n$ with $n=p,\ldots, 0$ such that the following relations are fulfilled:  
\begin{equation}
a_{n+1}+a_{n-1}+V_n\, a_n=0
\label{recurr}
\end{equation} 
Indeed, fixing $a_p=1$ from these recurrence relations, it follows that $a_{p-1}=-V_p$, $a_{p-2}=V_{p-1}V_p-1$ and so on, until a certain uniquely defined expression for $a_0$ (related to continued fractions).

Now consider the same recurrence relation (\ref{recurr}) but for negative $n=-p,-(p-1),\ldots,0$. 
Because, by the assumption (\ref{non_zero}) that $V_{-n}=-V_n$, one can choose solutions of (\ref{recurr}) for negative $n$ in the form
\begin{equation}
a_{-n}=(-1)^na_n
\end{equation}
with $a_n$ as above.

Due to the condition (\ref{zero}), the remaining equation (\ref{recurr}) with $n=0$ is also fulfilled, proving the existence of an eigenvector with zero eigenvalue and the vanishing of the determinant of the matrix considered.

\section{Effective action integral}\label{appI}

The purpose of this Appendix is to calculate the integral playing the role of the effective action for the whispering gallery mode in Section~\ref{continuous} for $x\to \infty$:
\begin{equation}
J(x)=\int_{m_1}^{m_2}\arccos(\rho_m(x))\mathrm{d}m\ .
\label{integral}
\end{equation}      
Here, $\rho_m(x)$ is defined in (\ref{rho}), $m_1=m_1(x)$ and $m_2=m_2(x)$ are solutions of equations
\begin{equation}
\rho_{m_1}(x)=-1\;,\;\;\rho_{m_2}(x)=1\ .
\end{equation} 
Their asymptotic values are given by (\ref{right_turning}) and  (\ref{left_turning}) correspondingly.

To find the asymptotics of the integral (\ref{integral}), we split it into 3 parts:
\begin{equation}
J(x)=J_1(x)+J_2(x)+J_3(x)\ .
\end{equation}
Here $J_1(x)$ denotes the integral over the region close to $m_1=m_p+1$ where $\rho_m(x)$ can well be approximated by the pole term
\begin{equation}
J_1(x)=\int_{m_p+1}^{M_1}\arccos \left (-\frac{1}{m-m_p}\right )\mathrm{d}m
\end{equation}
and $M_1-m_p\gg 1$.

$J_2(x)$ is the integral over the region far from the pole but close to $x$ where $\rho_m(x)$ is well described by the uniform approximation (\ref{airy})
\begin{equation}
J_2(x)=\int_{M_1}^{M_2}\arccos \left (-\frac{\mathrm{Ai}^{\prime}(y)}{v\mathrm{Ai}(y)} \right )\mathrm{d} m,
\label{J2}
\end{equation}
where $y=\frac{m-x}{v}$, $v=(x/2)^{1/3}$, and $M_2$ is chosen such that $x^{1/3}\ll M_2-x\ll x$.

Finally, $J_3(x)$ describes the integral over large $m$ where the approximation (\ref{semi}) is valid:
\begin{equation}
J_3(x)=\int_{M_2}^{m_2}\arccos \left (\sqrt{\frac{m^2}{x^2}-1}+\frac{x}{2(m^2-x^2)}\right )\mathrm{d}m\ .
\label{J3}
\end{equation} 
The integral $J_1(x)$ is straightforward, and one finds
\begin{eqnarray}
J_1(x)&=&\int_1^{M_1-m_p}(\pi-\arccos\frac{1}{u})\mathrm{d}u=
\pi(M_1-m_p-1)\\
&-&(M_1-m_p)\arccos \frac{1}{M_1-m_p}+
\ln \left ( M_1-m_p+\sqrt{(M_1-m_p)^2-1}\right )\ .\nonumber
\end{eqnarray}
In the limit $M_1-m_p\gg 1$, we obtain
\begin{equation}
J_1(x)=\frac{\pi}{2}(M_1-m_p)+\ln (M_1-m_p)+\ln 2 -\pi +1\ .
\end{equation}
As $v\equiv (x/2)^{1/3}\to \infty$ when $x\to \infty$ one can use the expansion $\arccos (z)\approx \pi/2-z $ in (\ref{J2}) 
\begin{equation}
\fl
J_2(x)=\int_{M_1}^{M_2}\left ( \frac{\pi}{2}+\frac{1}{v}\frac{\mathrm{Ai}^{\prime}}{\mathrm{Ai}}(\frac{m-x}{v})\right )\mathrm{d}m=
\frac{\pi}{2}(M_2-M_1)+\ln \mathrm{Ai}(\frac{M_2-x}{v})-\ln \mathrm{Ai}(\frac{M_1-x}{v})\ .
\end{equation} 
In the second Airy function one can use the approximation
\begin{equation}
\mathrm{Ai}(y)\approx \mathrm{Ai}^{\prime}(-\eta)(y+\eta),
\end{equation}
where as above $-\eta$ denotes the first zero of the Airy function, and in the first Airy function it is possible to use the asymptotic formula 
\begin{equation}
\mathrm{Ai}(y)\stackrel{y\to \infty}{\longrightarrow}\frac{1}{2\sqrt{\pi}}y^{-1/4}\exp \left (-\frac{2}{3}y^{3/2}\right )\ .
\end{equation} 
Thus one finds
\begin{equation}
\fl
J_2(x)\approx \frac{\pi}{2}(M_2-M_1)-\ln \left (\frac{M_1-m_p}{v}\right ) -\ln [2\sqrt{\pi} \mathrm{Ai}^{\prime}(-\eta)]
-\frac{1}{4}\ln \left (\frac{M_2-x}{v}\right )
-\frac{2}{3}\left (\frac{M_2-x}{v}\right )^{3/2}\ .
\end{equation}
In the integral (\ref{J3}) the second term is a small correction and 
\begin{equation}
J_3(x)\approx \int_{M_2}^{m_2}\left [ \arccos \sqrt{\frac{m^2}{x^2}-1}-\frac{x}{2\sqrt{2-m^2/x^2}(m^2-x^2)}\right ] \mathrm{d}m
\ .
\end{equation}
Substituting $m=xt$,
\begin{equation}
J_3(x)=x\left (\int_1^{\sqrt{2}}-\int_{1}^{M_2/x}\right )\arccos \sqrt{t^2-1}\, \mathrm{d}t 
-\frac{1}{2}\int_{M_2/x}^{\sqrt{2}}\frac{\mathrm{d}t}{\sqrt{2-t^2}(t^2-1)} \ .
\end{equation}
After simple transformations we obtain that $J_3(x)$ for $x\to \infty$ has the following asymptotics:
\begin{equation}
J_3(x)\approx I\,x-\frac{\pi}{2}(M_2-x)+\frac{2}{3}\sqrt{\frac{2}{x}}(M_2-x)^{3/2}+\frac{1}{4}\ln \left (\frac{M_2-x}{x}\right )\ 
\end{equation} 
where
\begin{equation}
I=\int_1^{\sqrt{2}}\arccos \sqrt{t^2-1}\, \mathrm{d}t=\sqrt{2}E\left (\frac{1}{\sqrt{2}}\right )-\frac{\pi}{2}\approx
.3393026
\label{elliptic}
\end{equation}
and $E(k)$ is the complete elliptic integral of the third kind,
\begin{equation}
E(k)=\int_0^1\frac{\sqrt{1-k^2t^2}}{\sqrt{1-t^2}}\mathrm{d}t\ .
\end{equation}
For later use we note that
\begin{equation}
\int_0^{\pi/2}\sqrt{1+\cos^2\phi}\mathrm{d}\phi=\sqrt{2}E\left (\frac{1}{\sqrt{2}}\right )\ .
\label{elliptic_2}
\end{equation}

Combining all the terms together, dependence on $M_1$ and $M_2$ indeed disappears as it should, and
\begin{equation}
J(x)\approx I\,x+\alpha x^{1/3}+\frac{1}{6}\ln x+\beta 
\label{full_integral}
\end{equation}
where 
\begin{equation}
\alpha=\frac{\pi}{2^{4/3}}\eta\approx 2.915016 ,
\end{equation}
\begin{equation}
\beta=-\frac{5}{12}\ln 2-\pi-\frac{1}{2}\ln \pi -\ln \mathrm{Ai}^{\prime}(-\eta)\approx -2.64782\ .
\end{equation}

\section{Shift of Bessel functions}\label{app_B}
Assume that $y$ is a root of the derivative of the Bessel function $J_p(x)$,
\begin{equation}
J_p^{\prime}(y)=0\ .
\end{equation} 
The purpose of this Appendix is to calculate $J_{p+k}(x)$ at the point $x=y$, taking into account two terms of the semiclassical expansion when $y\to \infty$ under the assumption that $p/x$ is finite but $k\ll x$. 
Our starting point is the asymptotic formula for the Bessel function, slightly more accurate than (\ref{bessel}), which can be found e.g.~in \cite{bateman},
\begin{equation}
J_p(x)=\sqrt{\frac{2}{\pi}}(x^2-p^2)^{-1/4}\left [ \cos \Phi_p(x) +\frac{b_1(x)}{\sqrt{x^2-p^2}} \sin \Phi_p(x)  +\mathcal{O}(x^{-2})\right ]
\end{equation} 
where $b_1$ depends only on $x/p$
\begin{equation}
b_1(x)=\frac{1}{8}-\frac{5}{24(1-x^2/p^2)}\ ,
\end{equation}
and $\phi_p(x)$ is the same as in (\ref{phase}).

Differentiating this expression over $x$, one finds that its root, $y$, obeys the equation
\begin{equation}
\sin \Phi_p(y)=\frac{1}{\sqrt{y^2-p^2}}\left [b_1(y)-\frac{y^2}{2(y^2-p^2)}\right ]\cos \Phi_p(y)\ . 
\label{definition}
\end{equation}
Using the semiclassical expansion of $\Phi_{p+k}(x)$, one gets  
\begin{equation}
\Phi_{p+k}\approx \Phi_p(x)-k\arccos \frac{p}{x}+\frac{k^2}{2\sqrt{x^2-p^2}}\ .
\end{equation}
Then
\begin{eqnarray}
J_{p+k}(x)&\approx & \sqrt{\frac{2}{\pi}}(x^2-(p+k)^2)^{-1/4}\left [ \cos \left (\Phi_p(x)-k\arccos \frac{p}{x}+\frac{k^2}{2\sqrt{x^2-p^2}}\right )\right .\nonumber \\
&+&\left . \frac{b_1(x)}{\sqrt{x^2-p^2}}\sin \left (\Phi_p(x)-k\arccos \frac{p}{x}\right ) \right ]\ .
\end{eqnarray}
Up to the second power of $1/x$, this equals
\begin{eqnarray}
J_{p+k}(x)&\approx &\sqrt{\frac{2}{\pi}}(x^2-p^2)^{-1/4}\left ( 1+\frac{pk}{2\sqrt{x^2-p^2}}\right ) 
\left [\cos \left (\Phi_p(x)-k\arccos \frac{p}{x}\right ) \right .\nonumber\\
&+& \left .\left (\frac{b_1}{\sqrt{x^2-p^2}}-\frac{k^2}{2\sqrt{x^2-p^2}}\right )\sin \left (\Phi_p(x)-k\arccos \frac{p}{x}\right )\right ]\ .
\end{eqnarray}
Expansion of the trigonometric functions and the definition of $y$ (\ref{definition}) leads to Eq.~(\ref{tchebichef_1}).

Using $2J_p^{\prime}=J_{p-1}-J_{p+1}$ and the recurrence relations for the Chebyshev polynomials \cite{bateman},
\begin{equation}
P_{k+1}(x)+P_{k-1}(x)=2xP_k(x)
\end{equation}
(where $P_k(x)$ stands for the Chebyshev polynomials of the first and the second kind), one may also obtain Eq.~(\ref{tchebichef_2}).

\section{Reflection coefficient for a periodic orbit}\label{app_A}

To find the total reflection coefficient for a given periodic orbit characterized by two integers $M$ and $N$, it is necessary to calculate the product (\ref{theproduct}) 
\begin{equation}
R(\phi)=\prod_{n=0}^{N-1}\frac{\sin \theta_{M,N}-\mathrm{i}\cos (\phi+2\theta_{M,N})}{\sin \theta_{M,N}+\mathrm{i}\cos (\phi+2\theta_{M,N})}
\end{equation}
where $\theta_{M,N}=\pi M/N$. 
Denoting $z=\mathrm{e}^{\mathrm{i}\phi}$, $\theta_{M,N}=\theta$, and $R(\phi)=R(z)$, one finds
\begin{equation}
(-1)^N R(z)=\prod_{n=0}^{N-1}\frac{z^2\mathrm{e}^{4\mathrm{i}\theta n}+2\mathrm{i}z \sin \theta \mathrm{e}^{2\mathrm{i}\theta n}+1}
{z^2\mathrm{e}^{4\mathrm{i}\theta n}-2\mathrm{i}z \sin \theta \mathrm{e}^{2\mathrm{i}\theta n}+1}\ .
\end{equation}
Expanding the quadratic polynomials in the numerator and the denominator into products of their roots, we obtain
\begin{equation}
(-1)^N R(z)=\prod_{n=0}^{N-1}\frac{(Y_n-\mathrm{i}u)(Y_n+\mathrm{i}v)}{(Y_n+\mathrm{i}u)(Y_n-\mathrm{i}v)}\ ,
\end{equation}
where $Y_n=z\mathrm{e}^{2\mathrm{i}\theta n}$, $u=1/v$, and 
\begin{equation}
v=\sqrt{1+\sin^2 \theta}+\sin{\theta}\ .
\end{equation} 
When $M$ and $N$ are coprime integers, all products in the above formula can be reduced to the following product:
\begin{equation}
\prod_{n=0}^{N-1}\left (t-\mathrm{e}^{2\pi \mathrm{i}/N n}\right )
\end{equation}
with a certain value of $t$. 
But the last product equals $t^N-1$, therefore
\begin{equation}
(-1)^N R(z)=\left [ \frac{(-\mathrm{i}z/u)^N-1}{(\mathrm{i}z/u)^N-1} \right ]
\left [ \frac{(\mathrm{i}z/v)^N-1}{(-\mathrm{i}z/v)^N-1}\right ]\ .
\end{equation}
Finally 
\begin{equation}
R(\phi)=\left \{ \begin{array}{c l}
1, & \mathrm{when }\; N=\mathrm{even}\\ 
\frac{\sinh \Lambda N-\mathrm{i}(-1)^{(N-1)/2}\cos \phi N}{\sinh \Lambda N+\mathrm{i}(-1)^{(N-1)/2}\cos \phi N}, 
&\mathrm{when }\; N=\mathrm{odd}\end{array}\right . 
\label{exact_reflection}
\end{equation}
where 
\begin{equation}
\Lambda=\ln \left (\sqrt{1+\sin^2 \pi M/N}+\sin \pi M/N \right )\ .
\end{equation}

\end{document}